\documentclass{aa}
\usepackage{graphicx}
\usepackage{subfigure}

\def\Ms{\mbox{${\rm M}_{\odot}$}}

\def\Msyr{\mbox{\Ms yr$^{-1}$}}

\def\lya{\mbox{${\rm Ly}{\alpha}$}}

\def\lumlya{\mbox{$L_{\rm{Ly}\alpha}$}}

\def\twwattm{\mbox{10$^{-21}$Wm$^{-2}$}}
\def\af{\mbox{$a_{\rm{flux}}$}} 

\def\al{\mbox{$a_{\rm{wave}}$}}
\def\ll{\mbox{${\lambda \lambda}$}}

\def\kms{\mbox{$\rm{km} s^{-1}$}}

\begin{document}

\title{\lya\ emission galaxies at a redshift of z $\approx$ 5.7 in the FORS Deep Field 
\thanks{Based on observations (Prop. ID: 071.A-0174(A) and 074.A-0237(A)) obtained at the ESO VLT
at Cerro Paranal, Chile.}}
\subtitle{}
\author{
C.~Tapken\inst{1,3}
        \and I.~Appenzeller\inst{1}     
        \and A.~Gabasch\inst{2,4}
        \and J.~Heidt\inst{1}
        \and U.~Hopp\inst{2,4}
        \and R.~Bender\inst{2,4}  
        \and D.~Mehlert\inst{3}
        \and S.~Noll\inst{4}
        \and S.~Seitz\inst{2,4}
        \and W.~Seifert\inst{1}
}
\institute{Landessternwarte Heidelberg-K\"onigstuhl, D-69117 Heidelberg, Germany
\and Universit\"ats-Sternwarte M\"unchen, Scheinerstr. 1, D-81679, M\"unchen, Germany
\and Max-Planck-Institut f\"ur Astronomie, K\"onigstuhl 17, D-69117 Heidelberg, Germany
\and Max-Planck-Institut f\"ur extraterrestrische Physik, Giessenbachstr., D-85741 Garching, Germany 
}

\offprints{C. Tapken, Heidelberg (\email{tapken@mpia.de})}
\date{received; accepted}

\abstract{We present the results of a search for  \lya\ emission galaxies 
at z~$\approx$~5.7 in the FORS Deep Field. The objective of this study is 
to improve the faint end of the luminosity function  of high-redshift \lya\
emitting galaxies and to derive properties of intrinsically faint \lya\
emission galaxies in the young universe. Using FORS2 at the ESO VLT and a set of special interference filters, we identified candidates for high-redshift
\lya\ galaxies. We then used FORS2 in spectroscopic mode to verify the
identifications and to study their spectral properties.
 The narrow-band photometry resulted in the detection of  15 likely 
\lya\ emission galaxies. Spectra with an adequate exposure time could be
obtained  for eight galaxies . In all these cases the presence of \lya\ emission  at $z$ = 5.7 was  confirmed spectroscopically. 
The line fluxes of the  15 candidates range between  3 $\times$ \twwattm\ and 
16 $\times$ \twwattm , which corresponds to star-formation rates not corrected
for dust between 1 and 5 \Msyr . The luminosity function derived for our
photometrically identified objects extends   the published luminosity functions of
intrinsically brighter \lya\ galaxies. With this technique the study of 
high-redshift \lya\ emission galaxies can be extended to  low 
intrinsic luminosities.

}

\maketitle
\titlerunning{LAEs at z $\approx\ $ 5.7 in the FDF}
\authorrunning{C.~Tapken et al.}
\section{Introduction}
The frequency and the physical parameters of high-redshift galaxies
provide important constraints on the formation and evolution of these
objects.   Strong  \lya\ emission allows  detection of
\lya\ emission galaxies (LAEs) at high redshift
from their excess in narrow-band filters (e.g. Kudritzki et
al. \cite{kudritzki2000};  Rhoads et
al. \cite{rhoads2000}; Steidel et al. \cite{steidel2000};  Fynbo et
al. \cite{fynbo2003}; Venemans et al. \cite{venemans2004}). Apart from drop-out -selected galaxies  
(e.g., Lehnert \& Bremer \cite{lehnert2003}; Bunker et al. \cite{bunker2004}), 
our knowledge  of the z $>$ 5 galaxies is based  on objects detected as
emitters of strong \lya\ radiation.  Well-defined gaps in the telluric
OH-bands allow  to detect LAEs  rather efficiently at very high redshifts  (Hu et
al. \cite{hu1998};  Taniguchi et al. \cite{taniguchi2003b}; Maier et
al. \cite{maier2003}; Ajiki et al. \cite{ajiki2003}; Rhoads et
al. \cite{rhoads2003}; Taniguchi et al. \cite{taniguchi2005}, Wang et
al. \cite{wang2005}; for a review see Taniguchi et al. \cite{taniguchi2003a}).
These gaps appear around $\lambda$ = 7110, 8160, 9210 \AA\ corresponding to
z$\approx$ 4.8, z$\approx$ 5.7   and z$\approx$ 6.6 for LAEs.   Particularly
successful have been observations  with the SuprimeCam of the Subaru
telescope, leading, e.g.,  to the discovery of a large-scale structure at
z$\approx$5.7 (Ouchi et al. \cite{ouchi2005}).  The properties of the
  LAEs have also been used to derive the luminosity function of LAEs  (Hu et al. \cite{hu2004}), the star-formation rate in the
early universe (Ajiki et al. \cite{ajiki2003}), and the epoch of re-ionization 
(Rhoads et al. \cite{rhoads2004}). However, so far, most detections of 
high-redshift LAEs have been based on relatively large area surveys sampling 
the intrinsically luminous objects. Only a few studies have been devoted
to faint LAEs at high redshift. Santos et al. (\cite{santos2004}) observed 
selected regions near intermediate-redshift clusters, finding eleven lensed 
low-luminosity LAEs between z = 2.2 and 5.6. Using a multi-slit technique, 
Martin et al. (\cite{martin2005}) find $\approx$ 20 
faint LAEs at z $\approx$ 5.7. 
 
Although bright \lya\ galaxies are relatively easy to detect, their
nature and physical structure is not understood well.
Since \lya\ is a resonance line of the most abundant element, the mean 
free path of \lya\ photons in the interstellar matter is short and 
the photons diffuse in physical and 
frequency space (Neufeld \cite{neufeld1990}). Hence, there 
is no straightforward correspondence between 
the properties of the \lya\ emission and the physical properties of the 
emitting galaxy.  Moreover, dust affects the \lya\ photons more 
strongly than the UV-continuum photons, which are not subject to resonance
scattering  (Charlot \& Fall \cite{charlot1993}). As dust absorption depends on the presence
of heavy elements, one may expect to also find among the LAEs 
metal-poor or primordial matter galaxies with 
a stellar population different from local objects  (Malhotra \&Rhoads \cite{malhotra2002}).
Because of the complex physics of the high-redshift \lya\ emission 
galaxies, their relationship with the better-studied continuum-selected 
high-redshift galaxies (LBGs) (see, e.g., Shapley et al. \cite{shapley2003}) 
is not understood well (Ouchi et al. \cite{ouchi2004}). A detailed derivation
and comparison of the 
luminosity functions of LAEs and LBGs could help to clarify
this relation. 
Furthermore, there are theoretical predictions that photoionization
heats and removes gas from the gravitational wells of galaxies
during the re-ionization period. This may suppressed the galaxy formation 
in low-mass halos (Barkana \& Loeb \cite{barkana1999}). A break in the luminosity function 
might be a tool for deriving the halo mass where the star-formation is 
suppressed (Santos et al. \cite{santos2004}) and other parameters of
the re-ionization epoch.  

We, therefore, carried out a search for intrinsically faint 
LAEs in the FORS Deep Field (FDF; Appenzeller et al. \cite{appenzeller2000};
 Heidt et al. \cite{heidt2003}; Noll et al. \cite{noll2004}).
Since the FDF covers an area of 7 ' $\times$ 7', only 
a few of the bright \lya\ galaxies detected in most of the surveys 
quoted above can be expected to be found in the FDF. However, 
by extrapolating the known part of the LF (Hu et al. 2004), we expected
(and verify with this paper) that a significant number of intrinsically
faint \lya\ emission galaxies are present in the FDF. A great advantage of the
FDF with respect to other surveys is the availability of very deep broadband
photometry and accurate photometric redshifts.  The B, R, and I broad-band
  images have a  limiting Vega magnitude (3$\sigma$ in 2'' diameter aperture)
  of 28.59,   27.99, and 27.21.
These data allow a reliable discrimination between 
\lya\ galaxies and other emission line objects. 
 
Throughout this paper we use Vega magnitudes and adopt  $\Omega_{\Lambda}$ = 0.7, $\Omega_M$ = 0.3, and H$_{0}$ = 70 km s$^{-1}$Mpc$^{-1}$. The probed redshift of 5.630 $< z <$ 5.803 results in a comoving volume of  1.5 $\times$ 10$^{4}$ Mpc$^3$.

\section{Photometric data}
\subsection{Observations and data reduction}
LAEs are identified by their redshifted \lya\ emission with narrow-band imaging. The choice of the band-width of the narrow-band filter is a tradeoff between the sensitivity and the search volume (Maier et al. \cite{maier2003}). Three narrow-band filters placed in one atmospheric gap are a reasonable choice in terms of sensitivity and search volume. Furthermore they have the advantage of determining the continuum of the object near the emission line (Klaus Meisenheimer, priv. comm.). For the present study we acquired three interference filters 
with bandwidths of about 60 \AA\ and central wavelengths of 8100 \AA , 8150
\AA , and 
8230 \AA\ with high effective transmissions ($\approx$ 93 \%)
for the FORS2 multimode instrument at the ESO VLT. The three slightly
overlapping filters fully cover the 8150 \AA\ window of the telluric OH emission.
Any object with emission lines in the wavelength range covered by the
filters is expected to show an excess emission in one or (in cases 
where the emission coincides with an overlap wavelength) two of the
filter bands.  After completion of
the present program, the filters are now publicly available. Their ESO  
designations are FILT\_810\_7+107, FILT\_815\_5+108, and FILT\_823\_6+109. In the following these designations will be abbreviated 
to F810, F815, and F823, respectively.

All photometric observations were carried out 
in August 2003 in visitor mode with FORS2 at the VLT-Yepun.  The detector 
was a mosaic of two 2k $\times$ 4k MIT CCDs used in the 
200kHz readout mode with low gain and 2$\times$2 binning.
Each frame was integrated for 1800 s. The total integration times and 
other information are listed in Table \ref{photo_nb}. The seeing 
varied  between 0$\farcs$5 and 0$\farcs$8. 

\begin{table}
\caption{Overview of the narrow-band observations. The ESO
  filter ID, their central wavelengths $\lambda _c$, their 
bandwidth $bw$ (FWHM), the total integration time and the 50 \% completeness 
limit are given.}
\label{photo_nb}
\vspace{2mm}
\centering
\begin{tabular}{c|c|c|c|c}
\hline
Filter & $\lambda _{c}$  & bw  & Exp. time  & 50 \% compl. \\
       & [\AA ]   & [\AA ] & [s] &  [mag] \\
\hline
FILT\_810\_7+107 & 8100 & 69 & 19800 & 25.56  \\
FILT\_815\_5+108 & 8150 & 53 & 21240 & 25.63  \\
FILT\_823\_6+109 & 8230 & 55 & 21240 & 25.67 \\
\hline
\end{tabular}
\end{table}

The data were reduced using the codes and procedures described by 
Heidt et al. (\cite{heidt2003}).  All
images were smoothed to a common seeing of 
FWHM=0$\farcs$78  and co-added signal-to-noise weighted.  The photometric 
calibration was done  using  standard stars with known spectra. The 
source detection was carried out on the co-added images using  SExtractor 
(Bertin \& Arnouts \cite{bertin1996}), with the same parameters as for 
the I-band images (for details see Heidt et al. \cite{heidt2003}). 
The number of significantly detected objects was 4303 in F810, 4519 in
F815, and 4675 in F823. SExtractor was also used to derive
total fluxes and magnitudes of the detected objects.  
 50\% completeness limits for point sources were calculated 
following  Snigula et al. (\cite{snigula2002}), using basic image and detection parameters. The completeness limits  are listed in 
Table \ref{photo_nb}.

The three filter bands avoid the stronger OH emission features 
still present in the 8150 \AA\ OH window by having most of these lines
coincide with the low-sensitivity wings of the narrow bands. Therefore,
the background is low, and lower emission-line-flux levels can be 
reached than is possible with a single narrow-band filter covering 
the 8150 \AA\ OH window. Other factors contributing to the high sensitivity 
of the present survey are the small filter band widths, the high 
sensitivity of the FORS2 CCD at 8150 \AA , and a better-than-average 
seeing during the observations. Moreover, all
narrow-band observations were made during the second half of the
observing nights, when (due to the nocturnal decay) the OH airglow 
is on average weaker. 

\subsection{Selection of the LAE candidates}
In order to find  LAE candidates among the objects that were
detected in the narrow band images 
we extracted in a first step all objects showing a statistically 
significant flux excess (relative to the other bands) in one or two of 
the three narrow band images.
Objects that had been detected in several  filter bands were regarded as having
a significant flux  excess if a flux difference larger than 
3 $\sigma $  was found.  As a next step we used  the deep broad-band
  images, the  FDF photometric catalog (Heidt et al. 2003), and the FDF
photometric redshift catalog (Gabasch et al. \cite{gabasch2004}) to eliminate
objects with $z<5.7$ where emission lines other than \lya\ coincided with our filter bands. 
In this way the great majority of the narrow-band excess objects could
be identified as foreground objects on the basis of their reliably lower
photometric (and in some cases spectroscopic) redshifts, blue flux, or too low
an (R-I).
The majority (85\%) of the narrow-band excess objects showed a small
narrow-band magnitude difference  of $<$ 1.2 mag. A significant fraction of
these objects have narrow-band excess caused by continuum features (e.g. we
found that cool stars can produce an excess flux in our narrow bands).  
Thus  an excess flux in a
narrow-band frame of 1.2 mag is required to select galaxies with a strong emission line. To select faint LAEs, we 
  used for the final catalog of LAEs candidates the deep I-band image as a
  reference.     About 50\%  of the excess objects with a narrow-band magnitude difference of $>$ 1.2 mag could be identified as [OII] emitters at $z=1.2$. Other narrow-band
flux excess objects could be explained by [O III] or Balmer line  emission.    
 Among the narrow-band excess objects that 
could not be eliminated as LAE candidates was only one object listed in the
Heidt et al. (2003) FDF photometric catalog. This object (FDF-2178, 
$z_{phot} = 5.74$) shows a marginally significant flux excess in 
the F823 filter and an  R-I $>$ 2.2. However, its flux in the F815 and F810 
bands is too high to be consistent with the expected drop shortward of
\lya . Hence we conclude that this object, although certainly a high 
redshift galaxy, is not a $z=5.7$ LAE candidate.

 The final catalog of LAEs candidates was selected in the following way.
First we selected all objects  showing a 3 $\sigma$
 flux excess (relative to the other bands) in one or two of 
the three narrow-band images. The narrow-band excess objects have to  satisfy 
the following criteria to be regarded as LAEs candidates: (1): NB $<$ 25.7 mag, (2):
I-NB $>$ 1.2 mag, (3): no detection in the B band.
 Bright objects (I $<$ 25.5 mag) had to fulfill the criterion (4a): R-I $>$
 1.5  mag and
faint objects the criterion (4b): no detection in the R band.  Fiffteen narrow-band
excess objects meet the criteria (1)-(4) and  are listed in Table 
\ref{phot_candidates}. In the following the individual objects 
are referred to by their sequence number in this table. The magnitudes 
given in Table \ref{phot_candidates} are narrow-band
magnitudes in the Vega system. Also included in Table 2 are
\lya\ equivalent widths (EW) derived (as described in Sect. 4.3)
from the photometric data and  photometric and (where available)
spectroscopic \lya\ line fluxes. The spectroscopic fluxes do not 
include corrections due to possible slit losses.
Some of the objects in the table are visible on the
FDF I-band and z-band images (however, in none of the bluer band
images) but have I-band fluxes below the 5 $\sigma $ limit. Therefore, none of objects in 
Table \ref{phot_candidates} is listed in the Heidt et al. (\cite{heidt2003}) FDF photometric catalog.

From a visual control of other crowded regions
of the narrow-band images, we regard it as unlikely that many  LAE
candidates have been missed by this selection criteria. Examples of the B, I,
F810, F815, and F823 images are presented in Fig. 1. An atlas of all objects in 
Table 2 in these filters is given in Fig. 9 (available in
electronic form only).

\begin{table*}
\caption{Photometric and spectroscopic properties of the  15 LAE
candidates. The limits of the narrow-band photometry correspond to the limiting magnitude 
(3$\sigma$ in 2'' diameter aperture).}
\label{phot_candidates}
\vspace{2mm}
\centering
\begin{tabular}{c|c|c|c|c|c|c|c|c|c}
\hline
Nr. & RA & DEC & mag$_{\rm{810}}$ & mag$_{\rm{815}}$ & mag$_{\rm{823}}$
& Flux$_{\rm{phot}}$        & EW [\AA ]      & $z_{\rm{spec}}$&
Flux$_{\rm{spec}}$       \\
    &    &     &                  &                  &                 
&  [10$^{-21}$ Wm$^{-2}$]   &                &                &
[10$^{-21}$ Wm$^{-2}$]    \\       
\hline
1& 1:05:50.36& -25:46:31.30 & $>26.35$  & $>26.46$ & 25.03 $\pm$ 0.11 &
5.41 $\pm$ 0.55 & $>$ 70.1 & 5.780  & 8.36 $\pm$  0.83 \\
2& 1:05:51.06& -25:43:21.50 &   25.28 $\pm$ 0.16 & 24.75 $\pm$ 0.09& 
$>26.48$ &5.02 $\pm$ 0.37 &$>$ 99.5 & -&- \\
3& 1:05:52.24& -25:47:10.24 & $>26.35$  & $>26.46$ &  25.12 $\pm$ 0.14 &
4.98 $\pm$ 0.64 & $>$ 61.8 & 5.770 &2.41$\pm$   0.49 \\
4& 1:05:52.45& -25:43:52.50 & $>26.35$  & $>26.46$  & 24.61 $\pm$ 0.10&7.97 $\pm$
0.73 &  $>$ 131.5 & -&- \\
5& 1:05:53.42& -25:44:26.42 & $>26.35$  & $>26.46$ & 24.88 $\pm$ 0.10 & 
6.22 $\pm$ 0.57& $>$ 86.9 & -&- \\
6& 1:05:54.94& -25:48:25.08 & $>26.35$   & $>26.46$ & 25.14 $\pm$ 0.15 &
4.89 $\pm$ 0.68 & $>$ 60.1&  5.787 & 8.17 $\pm$  1.02 \\
7& 1:05:58.45& -25:42:41.17 &  24.73 $\pm$ 0.10 & $>26.46$ & $>26.48$ &
9.02 $\pm$ 0.83 & $>$ 161.4&  -&- \\
8& 1:05:59.27 & -25:49:20.37 & 24.82 $\pm$ 0.11  & $>26.46$ & $>26.48$ &
8.30 $\pm$ 0.84 & $>$  138.0 &  5.627 & 4.79$\pm$   0.56 \\
9& 1:06:01.46& -25:44:25.45 & 25.16 $\pm$ 0.13  & $>26.46$ & $>26.48$ & 
6.07 $\pm$ 0.73&  $>$  80.7 & -&- \\
10& 1:06:06.05& -25:43:17.68 & $>26.35$  &  $>26.46$ & 25.69 $\pm$ 0.19 &
2.95 $\pm$ 0.52 & $>$ 28.4& -&- \\
11& 1:06:07.82& -25:48:49.86 &  $>26.35$  &  $>26.46$ &  25.07 $\pm$ 0.13 &
5.22 $\pm$ 0.62 &$>$  66.3 & 5.758  & 1.13 $\pm$  0.33 \\
12& 1:06:07.93& -25:43:17.47 &  $>26.35$  & 24.26 $\pm$ 0.05& $>26.48$ &
10.59 $\pm$ 0.49 &$>$  230.1 &  - &-\\
13& 1:06:08.88& -25:48:37.24 & $>26.35$  & 24.83 $\pm$ 0.11 &  
$>26.48$ &6.26 $\pm$ 0.63 & $>$88.4 &  5.679 &  4.37 $\pm$  0.50\\
14& 1:06:09.38& -25:48:20.32 & $>26.35$  & $>26.46$ & 25.00 $\pm$ 0.15 &
5.57 $\pm$ 0.77 &$>$ 73.1 & 5.764 & 4.05 $\pm$  0.54 \\
15& 1:06:16.29& -25:48:06.58 & 24.1 $\pm$ 0.05  & $>26.46$ & $>26.48$ & 
16.12 $\pm$ 0.74 & $>$ 872.4 & 5.640 & 11.99 $\pm$  0.72 \\

\hline 
\end{tabular}
\end{table*}

\begin{figure*}[htb]
\centering
\subfigure[]{\includegraphics[width=15cm]{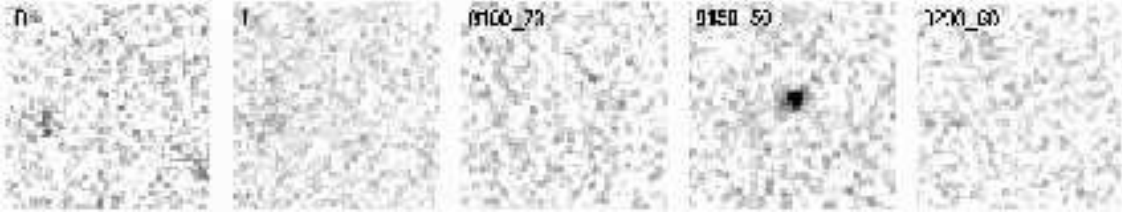}}
\subfigure[]{\includegraphics[width=15cm]{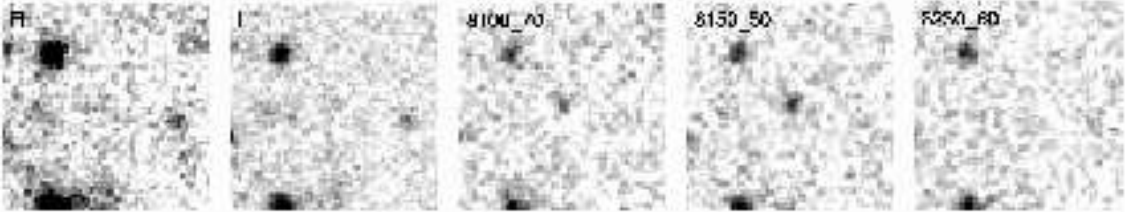}}
\caption {Bessel-R, Bessel-I broad-band images and F810, F815,
and F823  narrow-band images of 2 examples of our targets: (a) 
The second brightest candidate (No. 12) . (b)  Object with an
intermediate flux and visible in two narrow-band filters, although it 
was significantly detected  only in the F815 filter (No. 13). 
Each image shows a 10'' $\times$ 10'' area centered on the target.}
\label{nb_thumbnails}  
\end{figure*}

\section{Spectroscopic data}
\subsection{Observations and reduction}
Originally we planned to carry out a spectroscopic verification of all
LAE candidates in Table \ref{phot_candidates} using FORS2 in spectroscopic 
mode. Because of the high red efficiency of the FORS2 detector and the
FORS2 VPH gratings, a spectroscopic detection of the \lya\ flux (as estimated 
from the narrow-band data) of all our LAE candidates  can be achieved 
with a modest amount of observing time with this instrumentation.
An adequate amount of service observing time was granted for this program
by the ESO OPC in 2004. However, due to the 
pressure of higher priority programs only a small fraction of our
priority B program could actually be executed in November 2004. 
An application for additional time to complete the program was not successful.
 
For the spectroscopic observations obtained  
in November 2004, we used the holographic grism 1028z, covering 
the wavelength range 7800 - 9300 \AA\ at a resolution (with a 1'' slit width) of 
R $\approx$ 2700 ($\Delta v \approx$ 110 kms$^{-1}$). This resolution
allows a good subtraction of weak OH lines and a reliable 
identification of foreground [O II] emission objects (see Fig. 2i). To observe 
all 15 objects with an adequate S/N, three different MOS masks had been 
prepared, with the intention of  exposing each mask for three hours.  By 
including the  fainter objects in all three masks, while the brighter 
targets were included in only one mask, it was planned to detect all 
targets with an adequate SNR. For  
the reasons described above, only one mask 
could be exposed with total integration time of 2.7 h only.
 
The spectra were reduced using MIDAS and the routines developed by Noll 
et al. (\cite{noll2004}). 
The two-dimensional spectra were corrected with a dome flatfield and were  
wavelength-calibrated using the calibration spectra of gas discharge lamps. 
After the extraction of the  one-dimensional spectra, a flux calibration was
carried out 
using  spectra of standard stars observed during the same night.  The 
one-dimensional spectra were then co-added.  
Since the efficiency of VPH grisms varies with the angle of incidence 
and thus with the object's position in the telescope focal plane 
(see Tapken  \cite{tapken2005}), the fluxes were corrected for this 
using the sky-background. For 8000 - 8200 \AA , this effect was below 10 \%  
for most spectra.  

\subsection{Spectroscopic results}
\begin{figure*}
\centering
\subfigure[FDFLAE-1]{\includegraphics[clip=true,width=5.5cm]{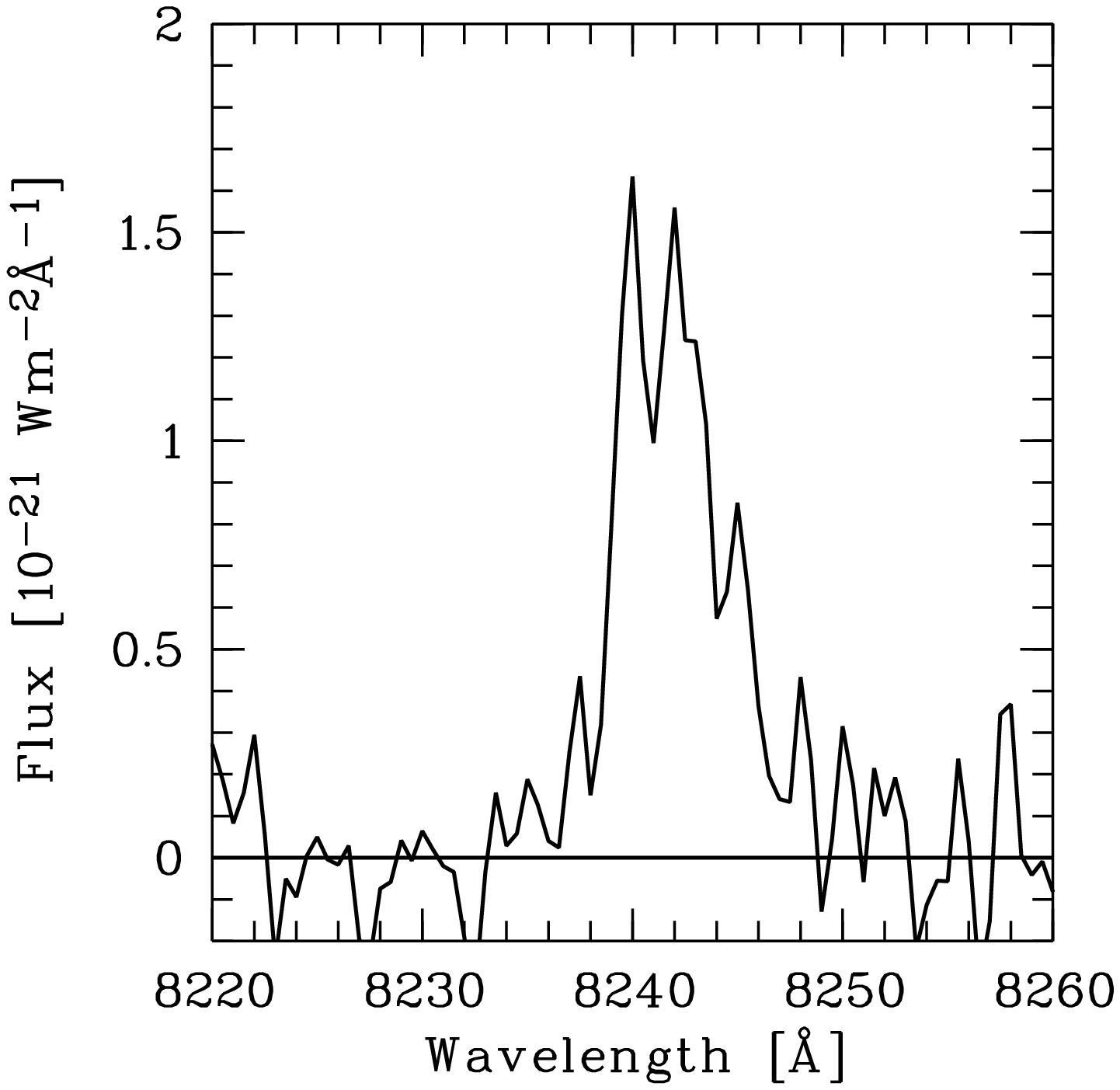}}
\subfigure[FDFLAE-3]{\includegraphics[clip=true,width=5.5cm]{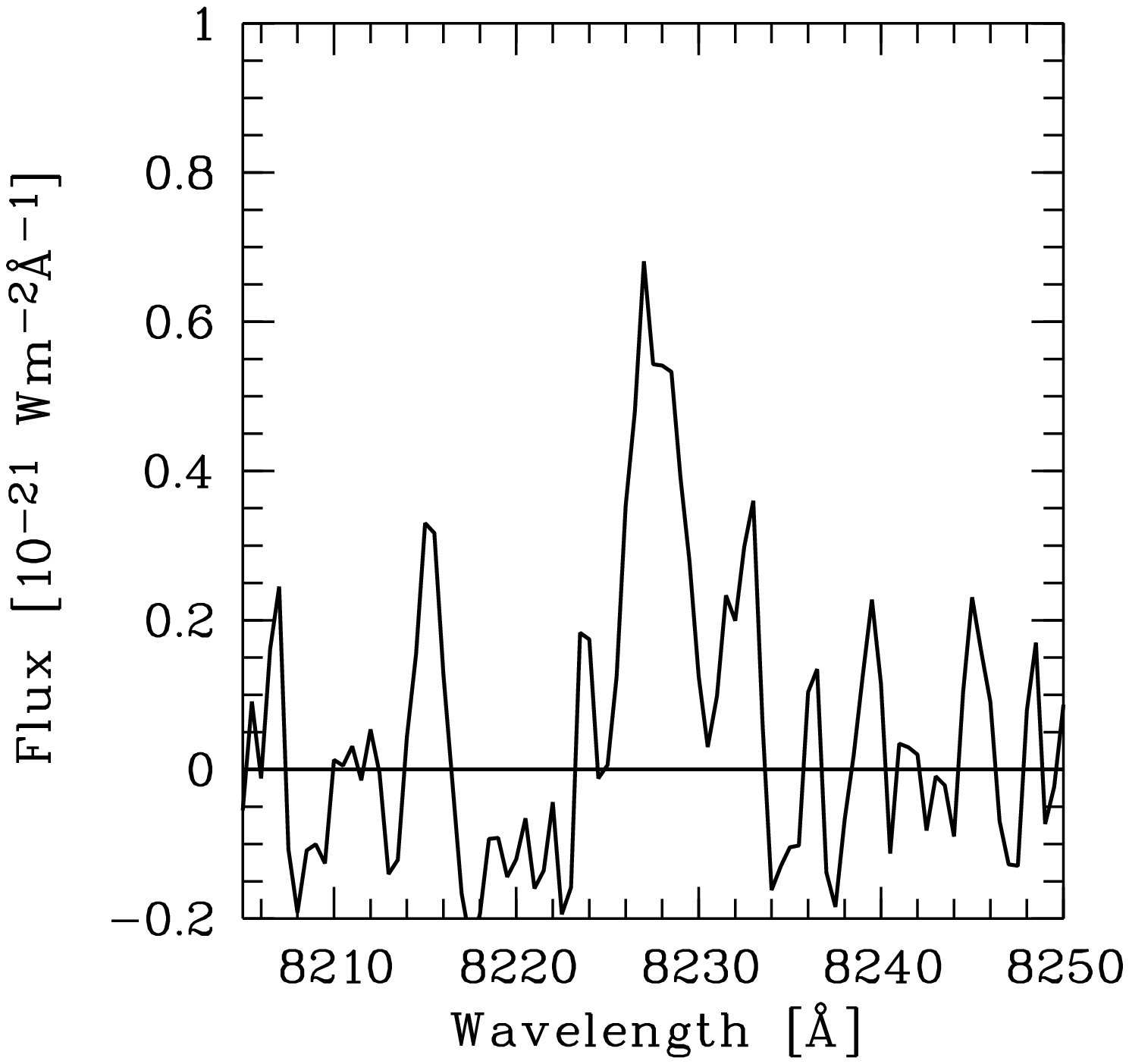}}
\subfigure[FDFLAE-6]{\includegraphics[clip=true,width=5.5cm]{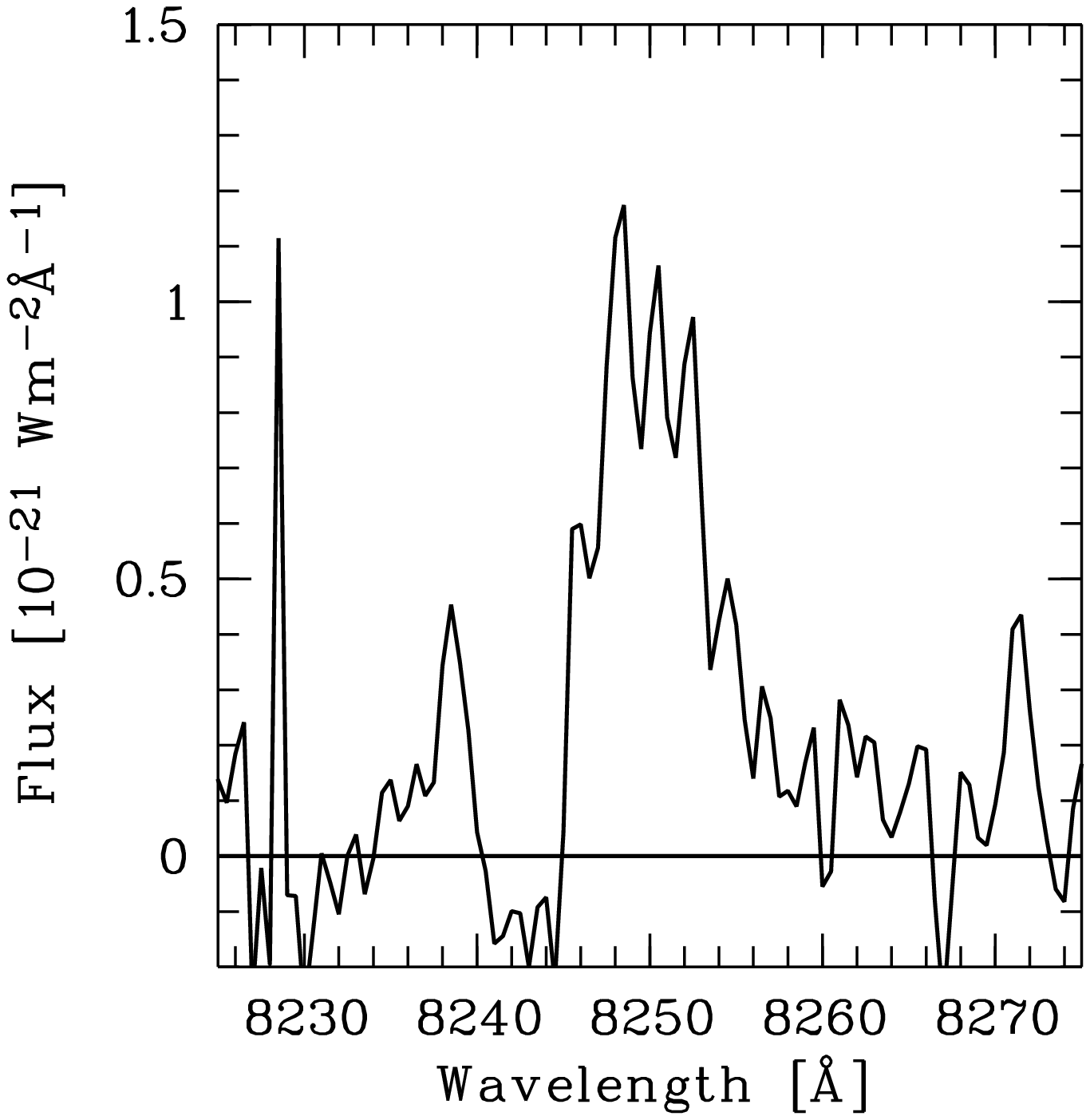}}
\subfigure[FDFLAE-8]{\includegraphics[clip=true,width=5.5cm]{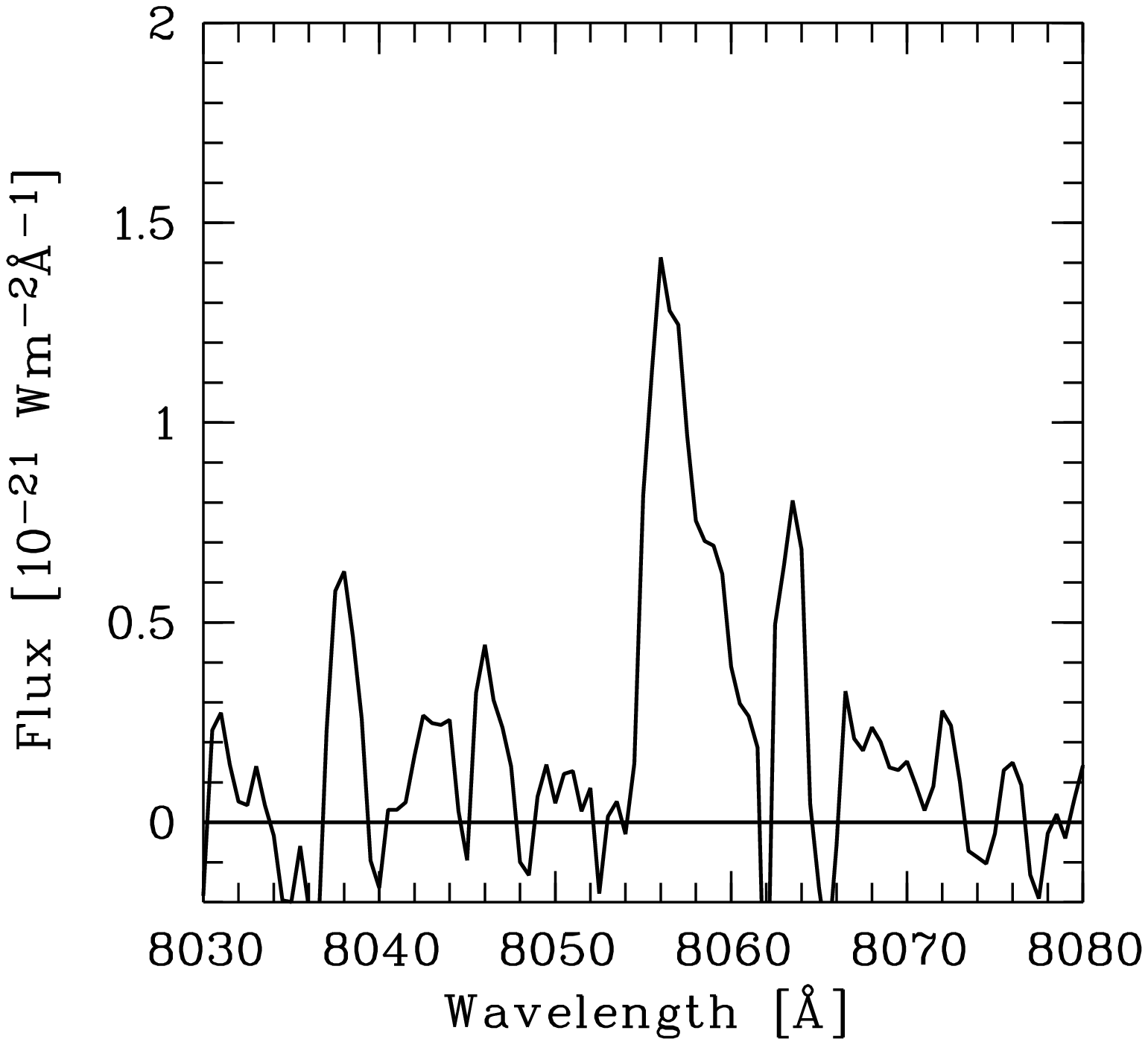}}
\subfigure[FDFLAE-11]{\includegraphics[clip=true,width=5.5cm]{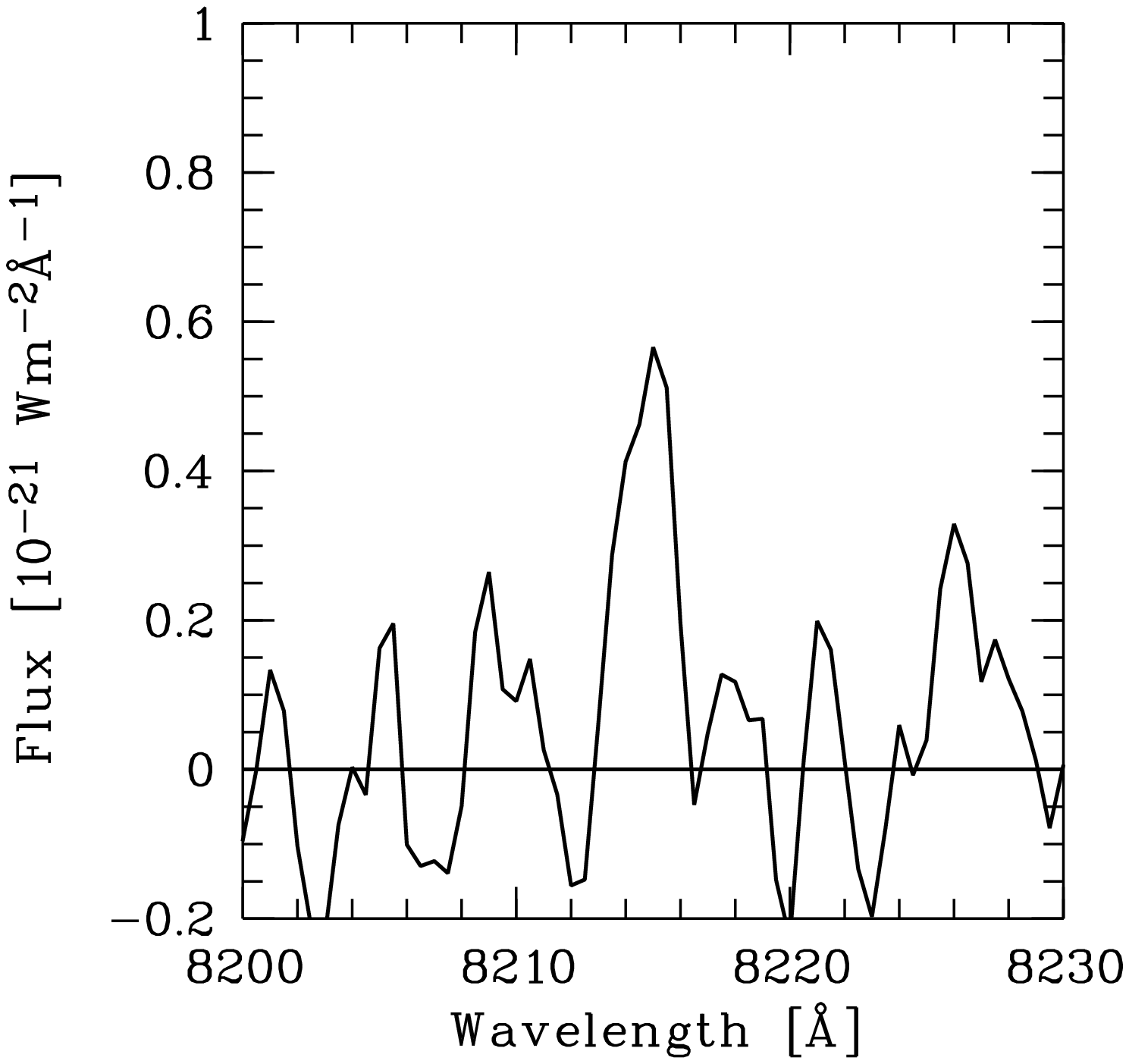}}
\subfigure[FDFLAE-13]{\includegraphics[clip=true,width=5.5cm]{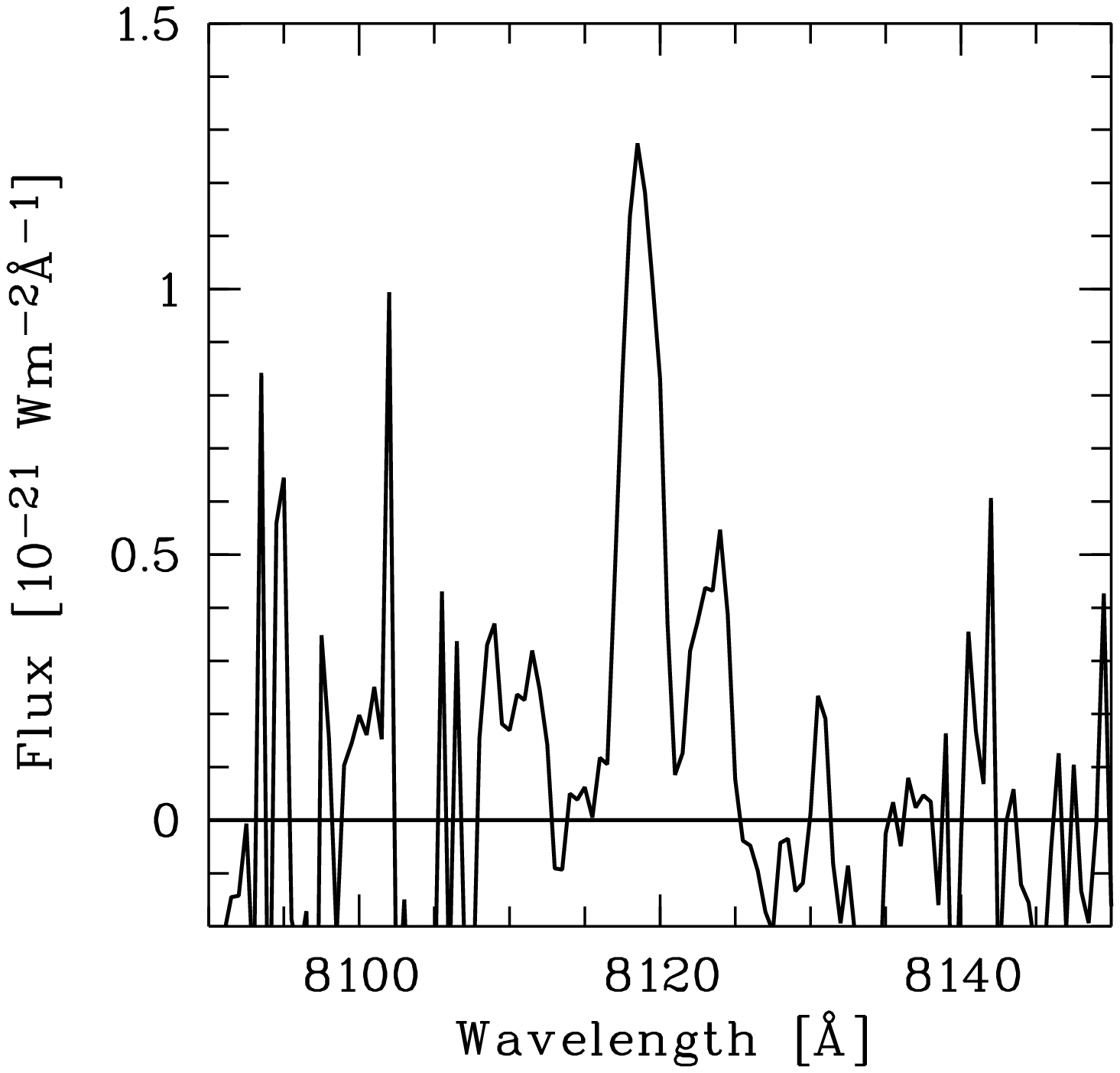}}
\subfigure[FDFLAE-14]{\includegraphics[clip=true,width=5.5cm]{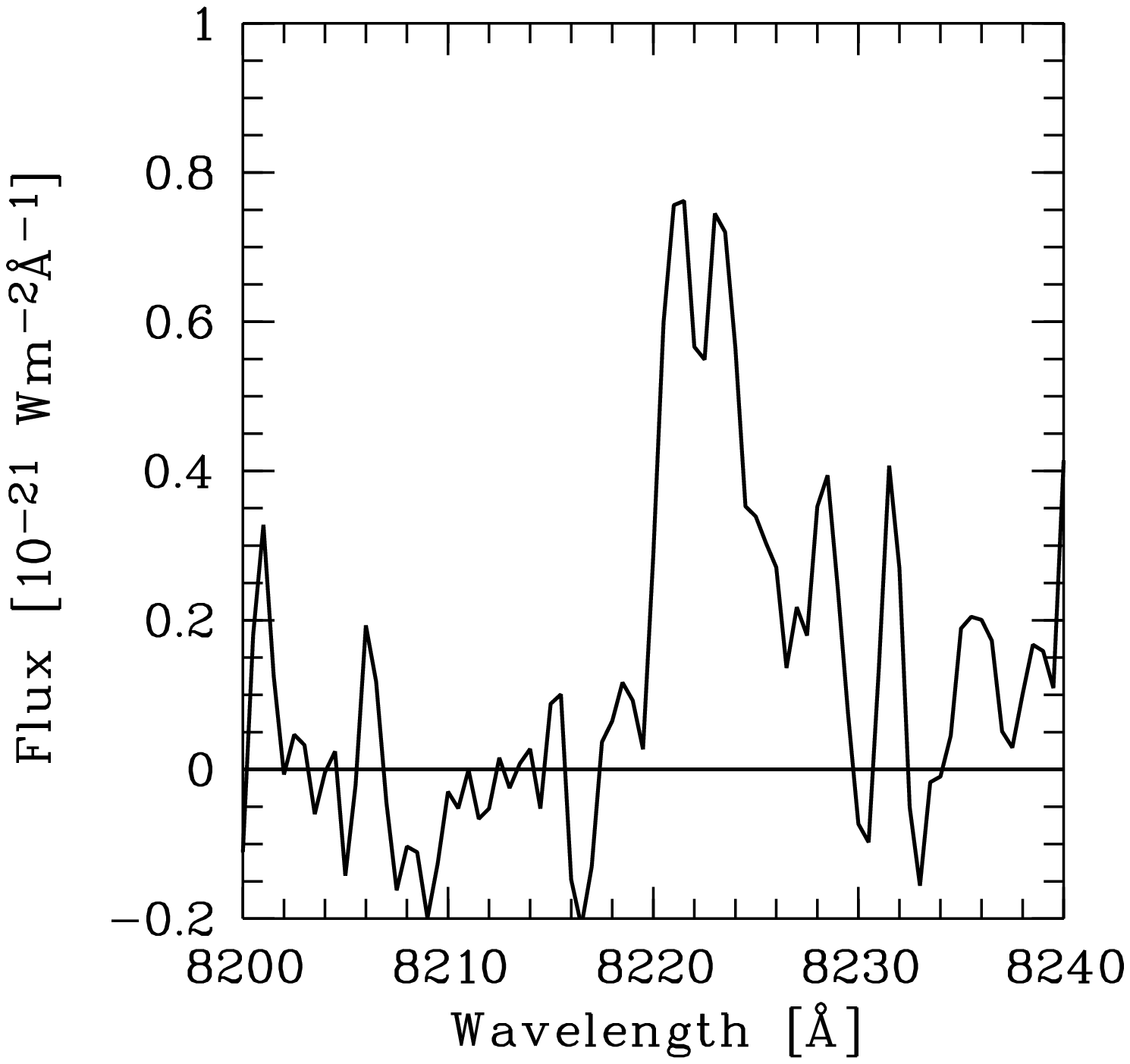}}
\subfigure[FDFLAE-15]{\includegraphics[clip=true,width=5.5cm]{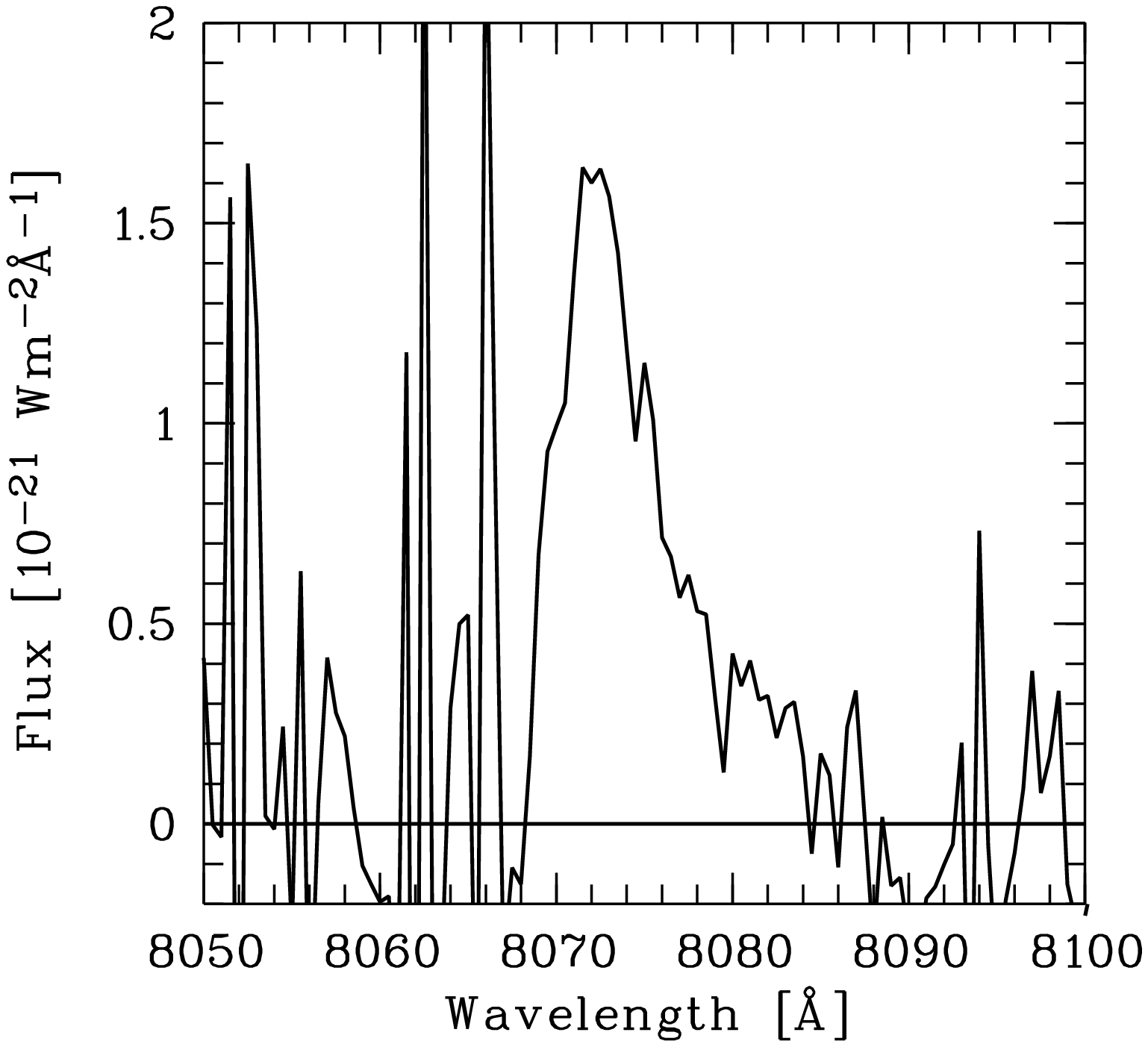}}
\subfigure[FDF-7757]{\includegraphics[clip=true,width=5.5cm]{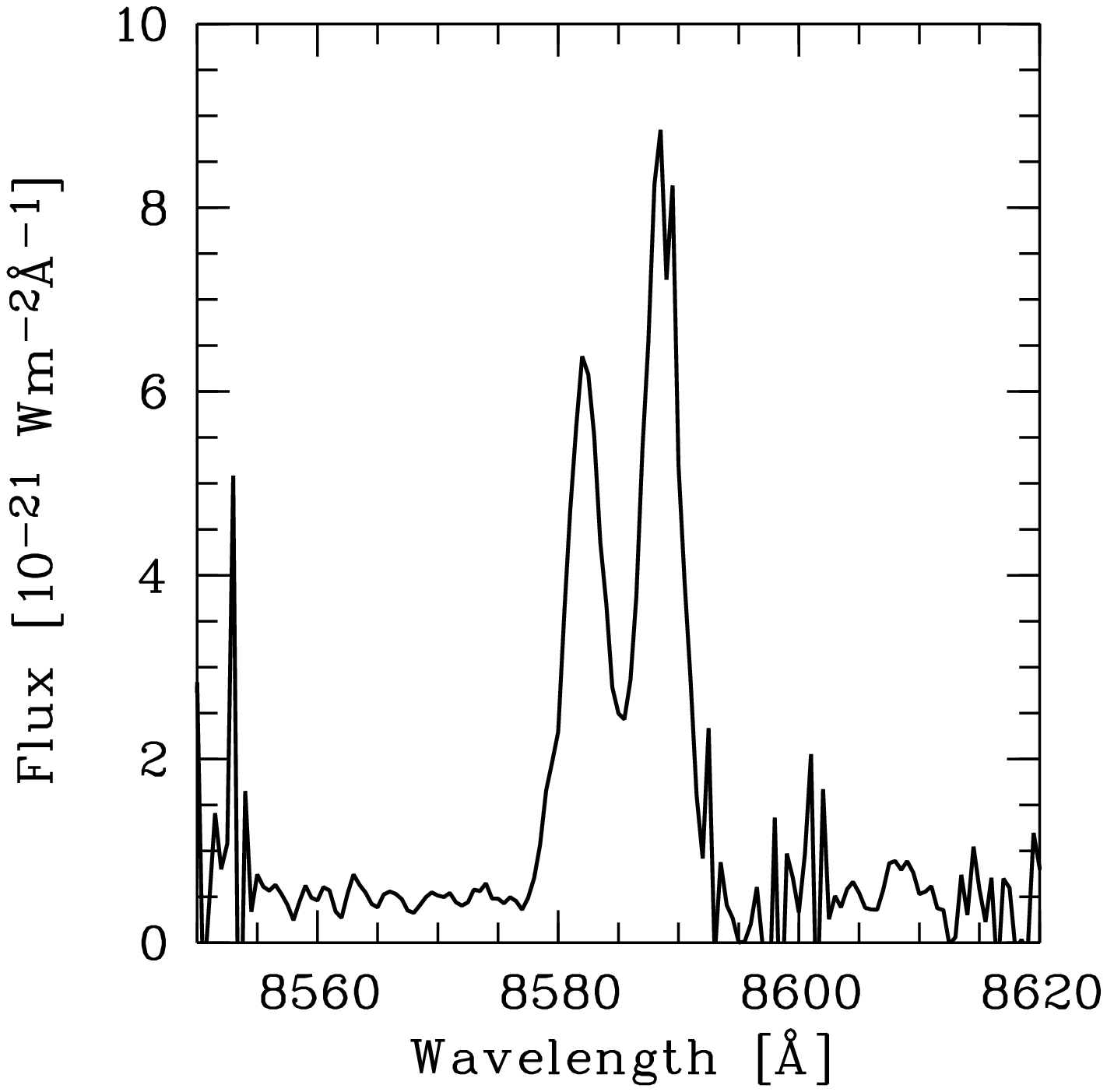}}

\caption {Observed emission line profiles of the LAE candidates. For comparison the [OII] 3727 doublet  of FDF-7757 at z$_{\rm{phot}}$ = 1.22 is also included.}
\label{line_profiles}  
\end{figure*}

The single MOS mask for which data could be obtained included 
 eight of the LAE candidates of Table \ref{phot_candidates} and several calibration and reference objects.
In spite of the short exposure time,  line emission from 
all of these eight LAE candidates was detected. In each case only one single 
emission line was observed. Figure \ref{line_profiles} shows the observed
line profiles.  Since all objects 
in Table 2
show only weak or no detectable continua, which could provide independent
information on the redshift, we verified that our spectra 
do not contain additional emission lines that were expected in the case that the observed 
features were due to redshifted [O III] or Balmer lines. Moreover, none
of the profiles has the structure expected for the [O II] 3727 \AA\ doublet.  
This doublet has a rest-frame separation of 2.8 \AA , leading to an observed 
doublet separation of $\approx$ 6 \AA , allowing for an easy separation 
at our resolution. As an example at the end of   
Fig. \ref{line_profiles} we include  the [O II] doublet of FDF-7757 
(z$_{\rm{phot}}$ = 1.22) observed during the same run. Only our LAE candidate
No. 13 shows an apparent (but not significant) doublet structure of
the expected width. However if this line was indeed [O II], the 
doublet ratio would  hardly be consistent with the theoretically allowed 
range of values (Osterbrock \cite{osterbrock1989}).

\section{Properties of the LAE candidates}

\subsection{Spatial and redshift distribution} 
\begin{figure*}[htb]
\centering
\includegraphics[angle=-90,bb = 80 80 480 480]{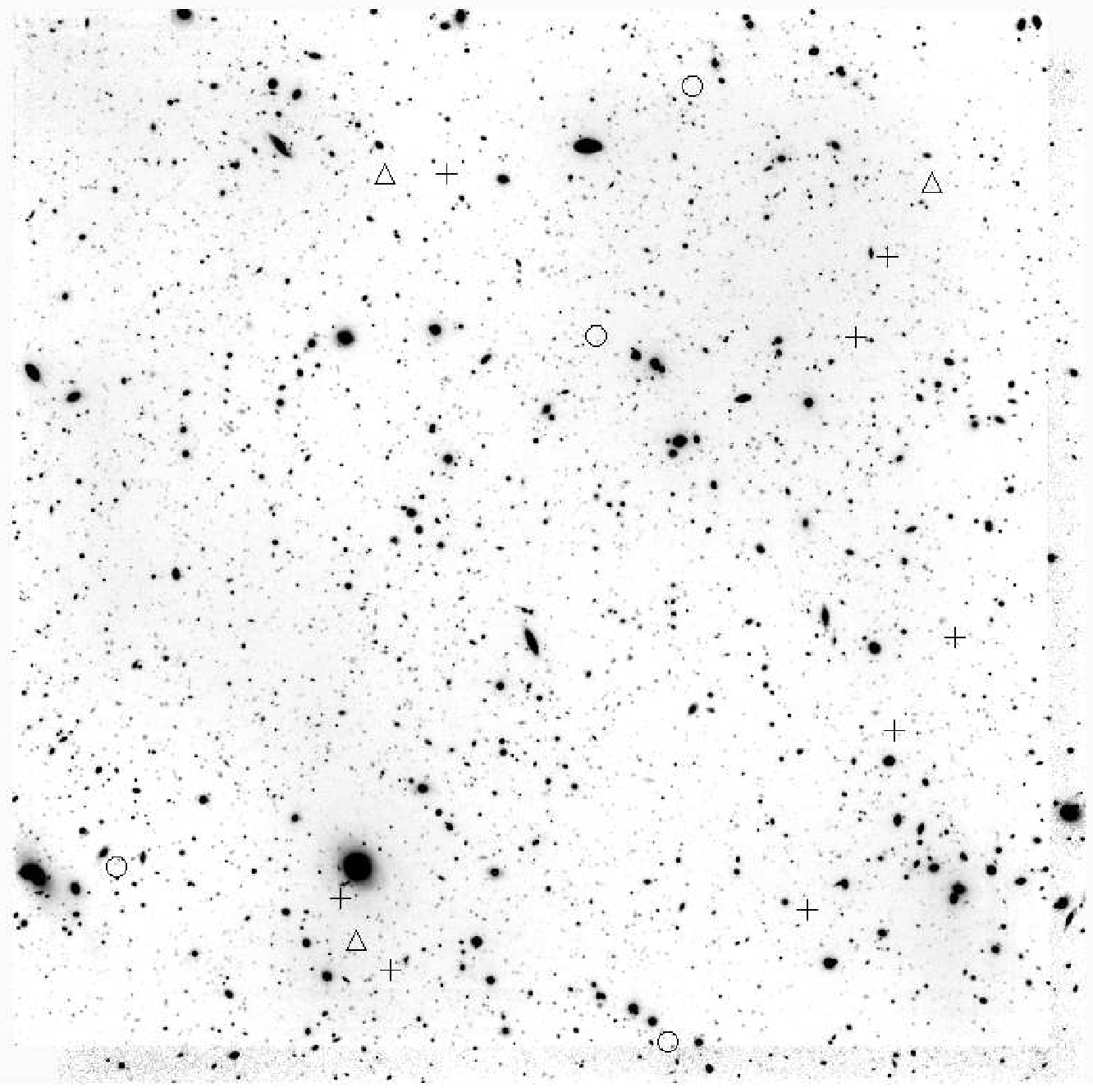}
\caption {The spatial distribution of the narrow-line excess objects in the FDF (shown is a deep I-band image). Circles correspond to objects found in the narrow-band nb810, triangles to objects detected in the narrow-band nb815, and crosses to objects found in the narrow-band nb823. North is up, east is left. The field is 7 $\times$ 7 arcminutes large, which corresponds to 2.5 Mpc $\times$ 2.5 Mpc (physical) or 17 Mpc $\times$ 17 Mpc (comoving).}
\label{fig_spatial_dis}  
\end{figure*}

Figure \ref{fig_spatial_dis} shows the spatial distribution of our LAE 
candidates. The objects are clustered at the edge of the field, while 
the center seems to be avoided. We verified that this distribution is not caused by reduction problems by comparing the limiting magnitudes in the de-voided area with the limiting magnitude in the area containing the objects. 
Our sample is too small to verify that
this distribution deviates significantly from a random distribution.
However, our result is consistent with the data of 
Hu et al. (\cite{hu2004}) covering of a $\approx$ 10 times larger volume, who find evidence
for large-scale structure of the bright LAEs distribution  and  Wang et
  al. (\cite{wang2005}), who find an overdensity of LAEs in the Chandra Deep
  Field South. 

\begin{figure}[htb]
\subfigure[]{\includegraphics[width=6cm,angle=-90]{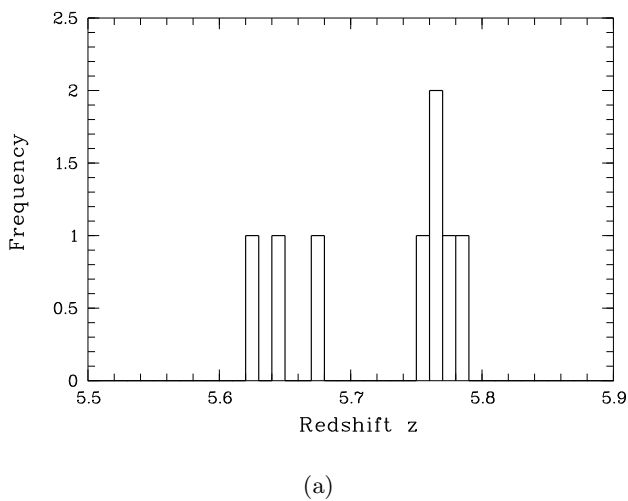}}
\caption {Distribution of redshifts of the eight LAE candidates 
observed spectroscopically.}
\label{fig_redshift_dis}  
\end{figure}

The distribution of spectroscopic  redshifts within the observational 
window is given in Fig. \ref{fig_redshift_dis}. A high fraction of F823 candidates were included in the mask observed, thus 
most of the redshifts  in Fig. \ref{fig_redshift_dis} are clustered at z $\approx$ 5.76. 

\subsection{Fluxes and star-formation rates}
\begin{figure}[htb]
\includegraphics[width=6cm,angle=-90]{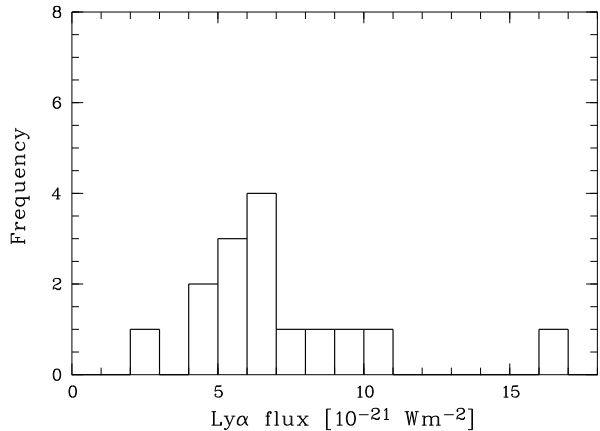}

\caption {The distribution of the photometric line fluxes of the 15 LAE candidates.}
\label{fig_flux_dis}  
\end{figure}

The line fluxes are derived using the narrow-band photometry. In
Fig. \ref{fig_flux_dis} the distribution of fluxes is shown. The fluxes range
from 3 $\times$ \twwattm\ to 16 $\times$ \twwattm . All fluxes are below the
flux limit of Hu et al. (\cite{hu2004}) of 20 $\times$ \twwattm .

\begin{figure}[htb]
\includegraphics[width=6cm,angle=-90]{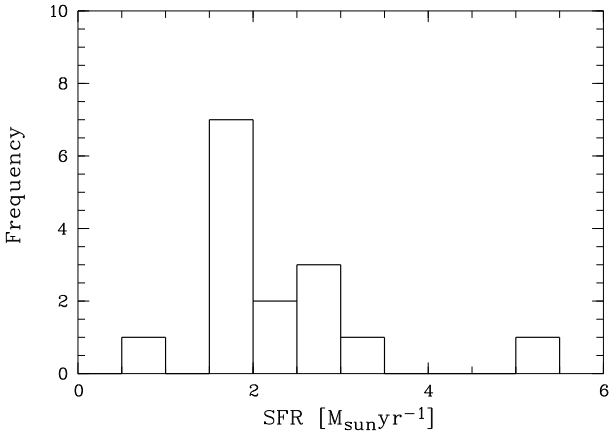}

\caption {The distribution of the star-formation rates derived from the 
photometric line fluxes.}
\label{fig_sfr_dis}  
\end{figure}

From the flux of the \lya\ line we computed star-formation rates. The 
line luminosities are derived  by assuming a cosmology model (
$\Omega_{\Lambda}$ = 0.7, $\Omega_M$ = 0.3, and H$_{0}$ = 70 km s$^{-1}$Mpc$^{-1}$) and 
isotropic emission. To derive the star-formation rates, the calibration of 
Kennicutt (\cite{kennicutt1998})\footnote{Kennicutt (\cite{kennicutt1998}) assumes a Salpeter initial mass function in the mass range of 0.1 to 100 \Ms, solar metalicity, and  a 100 Myr old starburst with continuous star-formation.}  was used and Case B recombination was 
assumed. In Fig. \ref{fig_sfr_dis} the distribution of the star-formation rates is given. 
The star-formation rates derived from the line fluxes are between 1 and 5 \Msyr . 
As pointed out above, part of the \lya\ may be lost by dust absorption. 
Moreover, the intergalactic medium probably absorbs a fraction of the \lya\ 
photons. Hence the observed line luminosity and thus the star-formation rates are only lower
limits. In addition, the conversion of  line luminosity into 
star-formation rates depends on assumptions of the 
starburst properties and the IMF.

\subsection{Equivalent widths}\label{sec_ew}
Using the I-band and the narrow-band photometric results, we  tried to
calculate \lya\ equivalent widths or lower limits for all our LAE candidates.
For this task we used the procedure described by Hu et al. (\cite{hu2004}). 
This
approach based on  the assumption of a rectangular filter shape of the
I-band and the assumption that the I-band flux represents the continuum. The
latter may be a poor approximation in our case.
We re-measured  the I-band magnitudes for all LAE candidates using 
aperture photometry. Although several objects are visible on the I-band
image, only four were marginally detected with 2 $<$ $\sigma$ $<$ 3.
For all objects, the three
sigma detection limit in a 2'' diameter aperture  was used as an upper limit for the continuum. In Table
\ref{phot_candidates} we list the resulting rest-frame equivalent widths
and lower limits. As shown by the table, all objects  have 
rest-frame equivalent widths $>$20 \AA .

\subsection{Line profiles}

All observed line widths of our LAEs are below 300 \kms  . This 
agrees with other spectroscopic studies of high-redshift LAEs 
(Dawson et al. \cite{dawson2004}). High-redshift LAEs have also be identified in the literature  on the basis of asymmetric line profiles, typically showing a sharp drop in the blue 
wing and an extended red wing  (Dawson et al. \cite{dawson2002}). Kunth et
al. also (\cite{kunth1998}) observed 
asymmetric \lya\ profiles also in nearby star-forming galaxies.  
Hu et al. (\cite{hu2004})  modeled the composite profile of 19 LAEs  with 
a Gauss emission profile and a Voigt absorption. They  attributed the 
absorption to the intergalactic medium. Most line profiles in Fig. \ref{line_profiles} also
show an asymmetry. 
 
To quantify the asymmetry we used the parameters introduced by Rhoads 
et al. (\cite{rhoads2003})  \af\ and \al .   \al\ gives the wavelength ratio  
and \af\ the flux ratio of the blue and red part of the \lya\ profile: 
\begin{equation}\label{eq_af}
a_{\rm{flux}}  = \frac{\int_{{\lambda}_{\rm{10,b}}}^{{\lambda}_{\rm{peak}}} f(\lambda) d\lambda}{  \int_{{\lambda}_{\rm{peak}}}^{{\lambda}_{\rm{10,r}}} f(\lambda) d\lambda} 
\end{equation}
and 
\begin{equation}\label{eq_aw}
a_{\rm{wave}} = \frac{(\lambda_{\rm{10,r}} - \lambda _{\rm{peak}})} {(\lambda _{\rm{peak}} - \lambda_{\rm{10,b}})}.
\end{equation}
Here, $\lambda _{\rm{peak}}$ is defined as the wavelength of the peak of the
emission. $\lambda_{\rm{10,b}}$ ($\lambda_{\rm{10,r}}$) is defined as the
wavelength where the flux reaches 10 \% of his peak flux level on the blue (red) side.

\begin{figure}[htb]
\resizebox{\hsize}{!}{\includegraphics[clip=true]{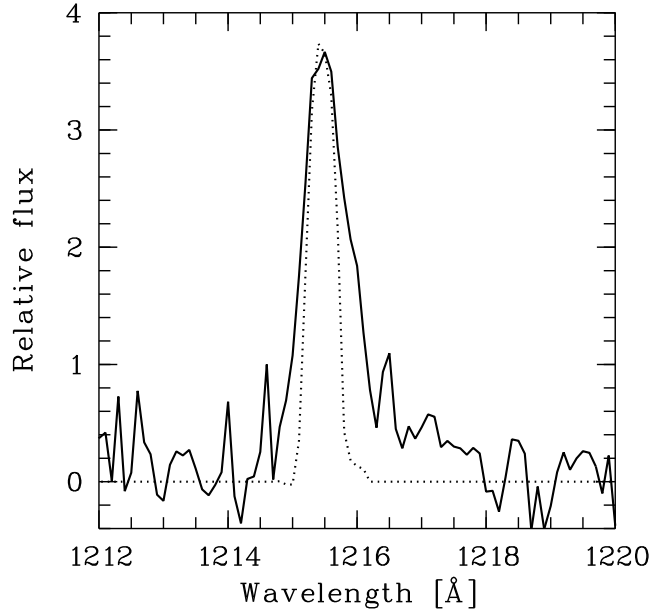}}
\caption {Composite spectra of  the eight emission line objects. The dotted line is the instrumental profile, derived from the sky lines.}
\label{fig_comp}  
\end{figure}

For six lines we could measure  the values of \af\ and \al . Except
for FDFLAE-13, all objects show that \af\ and \al\ are greater than 1, i.e., are
extended to the red in good agreement with the results of  Dawson et
al. (\cite{dawson2004}), who mainly analysed galaxies with $z > 4$ .  
Moreover, Dawson et al. (\cite{dawson2004}) found for 28 galaxies at 
$z \approx 1$  asymmetry parameters of the [OII] \ll 3726.2, 3728.9  line 
of \al\ and \af\  $\approx 1$. In our sample FDFLAE-15 has the strongest 
asymmetry 
with \af\ = 3.91 $\pm$ 0.57 and \al\ = 1.64 $\pm$  0.26. 
A composite spectrum formed by simply adding all our individual profiles
is shown in Fig. \ref{fig_comp}. Obviously this composite 
also shows a marked  asymmetry in the expected sense.

From the discussion above, we conclude that at least our LAE candidates  
with a spectroscopic confirmation are indeed LAEs at $z$ = 5.7.
The high success rate of our spectroscopic verifications
 suggests that at least the great majority of our all LAE candidates are  LAEs
 at $z$ $\approx$ 5.7.

\subsection{The luminosity function}
\begin{figure}
\resizebox{\hsize}{!}{\includegraphics[angle=-90]{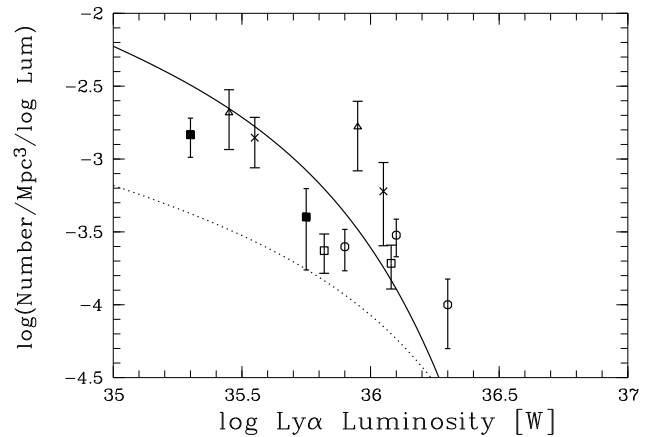}}
\caption {Luminosity function derived from the photometric data (filled
  squares). Two surveys at a redshift of $z$ = 5.7 have been included: circles
  correspond to Hu et al. (\cite{hu2004}) and open squares to Ajiki et
  al. (\cite{ajiki2003}). Two surveys at a redshift of z$\approx$ 3.2 are also
  included: triangles correspond to Kudritzki et al. (\cite{kudritzki2000}),
  crosses to Cowie \& Hu (\cite{cowie1998}).  Moreover, a Schechter function
  (Breukelen et al. \cite{breukelen2005}) for \lya\ emitters with 2.3 $<$ $z$
  $<$ 4.6  (solid line) and a Schechter function (Malhotra \& Rhoads
  \cite{malhotra2004}) for LAEs at z$\approx$ 5.7 (dotted line) are plotted.
   The error bars denote statistical mean errors. }
\label{lumfunc}  
\end{figure}

Using the observed density of our LAE candidates in the FDF and assuming 
(for the reasons given above) that all these objects are LAEs, we can
use the data of Table 2 to derive the surface and space density 
of the low-luminosity LAEs as a function of the luminosity (i.e. the LF). 
In order to avoid uncertain incompleteness 
corrections, we used for
this purpose only those objects in Table 2 that have narrow-band
magnitudes brighter than the 50 \% completeness limit (14 of the  15 LAEs). The results are
given in Fig. \ref{lumfunc},  together with the 
published luminosity function of Hu et al. (\cite{hu2004}) and Ajiki et al. (\cite{ajiki2003})
. Hu et al. (\cite{hu2004}) used deep narrow-band imaging of the SSA22 field  to search for 19 LAEs 
at a redshift of $z$=5.7. They detected 26 candidates down to a limited flux of 20  
$\times$ \twwattm . Spectroscopic observations showed a high succes rate.  Ajiki et al. (\cite{ajiki2003}) found 20 LAE
candidates at a redshift of $z$=5.7 with the Suprime-Cam.
As shown by the figure, our luminosity function is consistent with a
continuation of the Hu et al. and Ajiki et al.  luminosity function towards lower 
luminosities.  Malhotra \& Rhoads (\cite{malhotra2004}) assembled data from
several narrow-band LAEs surveys at a redshift of z $\approx$ 5.7 and derived
a Schechter function ($\alpha$ = - 1.5, L$_{\star}$ =
10$^{36}$ W, $\Phi_{\star}$ = 10$^{-4}$ Mpc$^{-3}$)  included in
Fig. \ref{lumfunc}. We find more LAE candidates than expected
on the basis of this luminosity function.

We also included in Fig. \ref{lumfunc} the luminosity function from Kudritzki 
et al. (\cite{kudritzki2000})   and Cowie \& Hu (\cite{cowie1998}).  Kudritzki et al. (\cite{kudritzki2000})   and Cowie \& Hu (\cite{cowie1998}) 
used the narrow-band technique to detect LAEs at a redshift of z=3.1
(3.4). Their  results are largely consistent with the higher redshift luminosity function. 
Moreover, we included the Schechter function for \lya\ emitters with 2.3 $<$ 
$z$ $<$ 4.6 (mean $z$ = 3.18)  derived by van Breukelen et al. (\cite{breukelen2005}). Van Breukelen et al. (\cite{breukelen2005}) detected 14 LAEs
using VIMOS IFU spectroscopy. Their Schechter function is consistent with our results. 

We also compared our number counts with the prediction of Thommes \&
Meisenheimer (\cite{thommes2005}), who compute the expected surface density of \lya\ emitting galaxies on the assumption that LAEs are the progenitors of  today's elliptical galaxies and bulges of spiral galaxies.
One of the main assumptions of this model is a short \lya\ bright phase of the
star-forming galaxies. This short duration of the \lya\ bright phase is
explained 
by the rapid formation of dust in this galaxies. The basic model  of Thommes \& Meisenheimer (\cite{thommes2005}) predicts a  number of $\approx$ 20 LAEs with \lumlya\ $\ge$ 10$^{35}$ W in the FORS Deep Field at a redshift of z$\approx$ 5.7. Given the large uncertainity, this  agrees with the observed number of 15.   
 
\section{Conclusions}
Our survey has shown that FORS2, the ESO VLT, and our special
filter set make it possible to extend groundbased searches for
\lya\ emission galaxies at $z = 5.7$ to significantly lower luminosities. 
Although the lack of observing time did not allow a full spectroscopic 
verification of all objects detected by our narrow-band filter survey,
the available spectroscopic results indicate a high success rate of this
set-up. The observed faint LAEs show  very similar properties to those
observed at higher luminosities. A luminosity function constructed on the
assumption that all our photometrically detected emission objects
are LAEs at $z$ = 5.7 is in good agreement with the LF published by 
Hu et al. (\cite{hu2004}) and extends the Hu et al. luminosity function 
towards lower luminosities.

\begin{acknowledgements}
We would like to thank the staff of the Paranal Observatory for
carrying out the spectroscopic service observations.
This research has been supported by the German Science Foundation DFG 
(SFB 439). We thank the referee for valuable comments.
\end{acknowledgements}

\end{document}